\documentclass[twocolumn,showpacs,prl]{revtex4}
\usepackage{amssymb}

\usepackage{amsmath}
\usepackage{epsf}
\usepackage{graphicx}
\usepackage{dcolumn}
\usepackage{bm}
\usepackage{amsmath}

\newcommand{\beq}{\begin{equation}}
\newcommand{\eeq}{\end{equation}}
\newcommand{\bea}{\begin{eqnarray}}
\newcommand{\eea}{\end{eqnarray}}
\newcommand{\bwt}{\begin{widetext}}
\newcommand{\ewt}{\end{widetext}}
\newcommand{\otau}{1/\tau_{\mathrm{ee}}}

\newcommand{\bk}{{\mathbf k}}
\newcommand{\bp}{{\mathbf p}}
\newcommand{\ti}{\tau_{\mathrm{i}}}
\newcommand{\bb}{{\mathbf b}}
\newcommand{\bq}{{\mathbf q}}
\newcommand{\bv}{{\mathbf v}}
\begin{document}

\title{Resistivity of a non-Galilean--invariant Fermi Liquid\\ near Pomeranchuk Quantum Criticality}
\author{Dmitrii L. Maslov\;$^{a}$, Vladimir I. Yudson\;$^{b}$, and Andrey V. Chubukov\;$^{c}$}
\date{\today}

\affiliation{
$^{a}$Department of Physics, University of
Florida, P. O. Box 118440, Gainesville, FL
32611-8440\\
 $^b$Institute for
Spectroscopy, Russian Academy of Sciences, Troitsk, Moscow region,
142190, Russia\\
$^c$Department of Physics, University of
Wisconsin-Madison, 1150 Univ. Ave., Madison, WI 53706-1390}

\begin{abstract}
We analyze the effect of the electron-electron interaction on the resistivity
 of a metal near a Pomeranchuk quantum 
 phase transition (QPT).
We show that Umklapp processes are not effective near a QPT, and one must
 consider  both interactions 
 and  disorder to obtain finite and $T$ dependent resistivity.  By power counting, the correction to the residual resistivity at low $T$ scales as $AT^{(D+2)/3}$ near a $Z=3$ QPT. We show, however,  that  $A=0$ for a simply connected,
  convex Fermi surface in 2D,  
  due to hidden integrability of the electron motion. We argue that $A >0$ in a two-band ($s-d$) model   and propose this model 
as an explanation for the observed $T^{(D+2)/3}$ behavior.
\end{abstract}

\maketitle
A $T^2$ scaling
   of the resistivity $\rho$  is the main signature of the Fermi liquid (FL) behavior in metals. Although this 
   scaling 
 is usually associated with the $T^2$ 
 behavior  of the quasiparticle scattering rate, $1/\tau_{\mathrm{ee}}$, the relation between $\otau$ and $\rho$ is 
 not
 straightforward
 because one has to specify a momentum relaxation mechanism. 
 For example, even though $\otau\propto T^2$ in a Galilean-invariant FL (GIFL), its resistivity is zero
 (although the heat conductivity and viscosity are finite). 
In clean systems and at low $T$ (when scattering on phonons can be neglected),
the mechanism of momentum relaxation is Umklapp electron-electron  ({\em ee}) scattering \cite{peierls:29,landau:36},
 which conserves the quasimomentum
up to a reciprocal lattice vector: $ {\bf k}+{\bf p}={\bf k}'+{\bf p}'+{\bf b}$. An Umklapp process is allowed if  the electron momenta 
 ${\bf k}$ and ${\bf p}$, as well as  the 
 momentum transfer ${\bf q} = {\bf k} - {\bf k}'$, are all
 of  order ${\bf b}$; these two
  conditions are usually satisfied in conventional metals. If this is the case, Umklapps occur at a rate comparable to $\otau$, and $\rho \propto \otau$. 

Even a conventional metal, however, can be tuned to a 
 Pomeranchuk-type 
quantum phase transition (QPT) lowering the symmetry of the 
Fermi surface (FS).
\begin{figure}[t]
\includegraphics[width=0.4\textwidth]{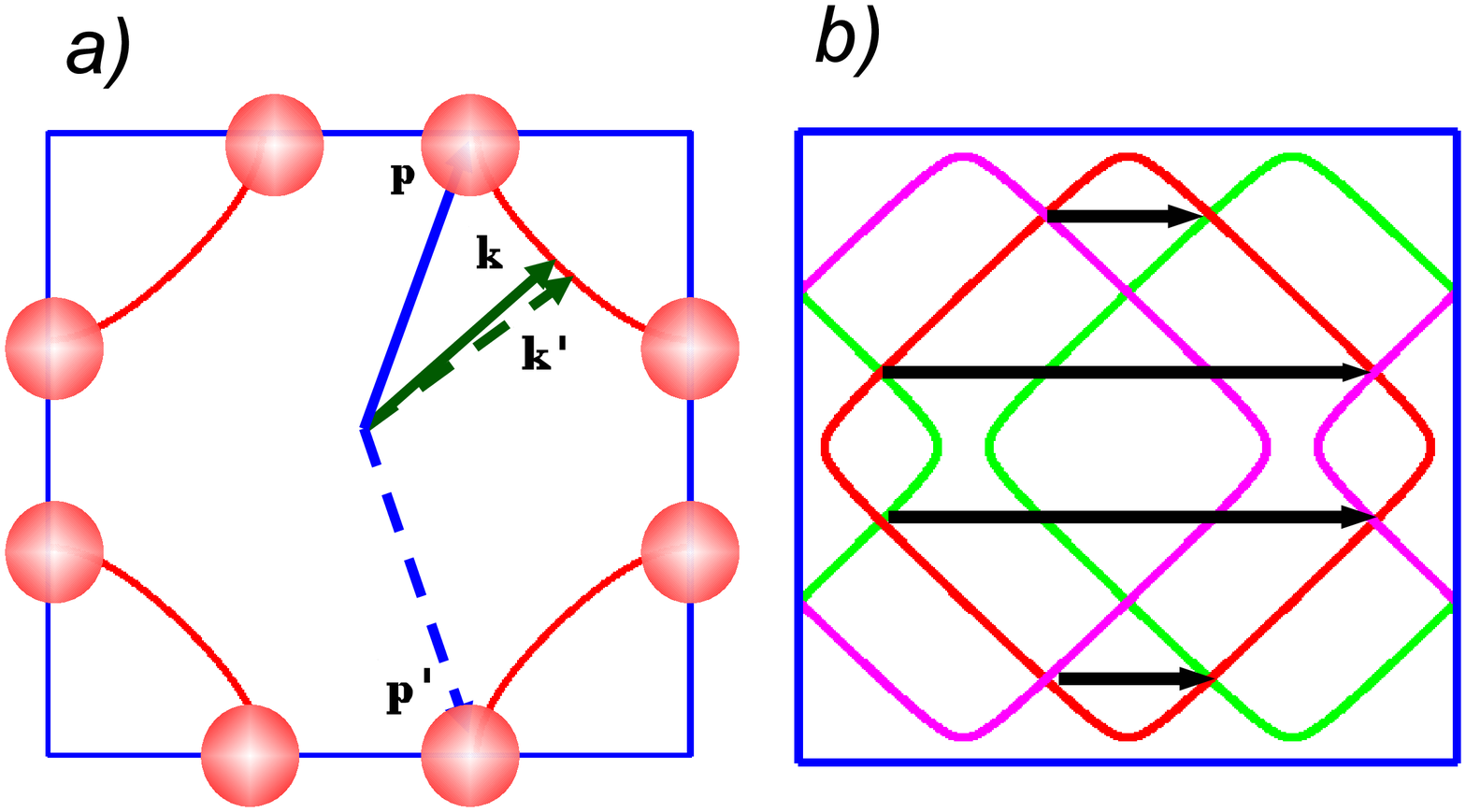}
\caption{(COLOR ONLINE) {\em a}) Umklapp process for a long-range electron-electron interaction. {\em b})  
Umklapp processes for large $q$. The original FS is shown in red. }
\label{fig:fig1}
\end{figure}
A Pomeranchuk QPT is the $q=0$ instability, manifested by the divergence of long  wavelength fluctuations of the order parameter \cite{pomeranchuk:58}. The FL near a Pomeranchuk QPT differs from that in a conventional metal in that the effective {\em ee} interaction is  of a long-range. In the Hertz-Millis model,
\beq
U_{\mathrm{eff}} (\bq,\omega)=\frac{U_0} {q^2+\xi^{-2}-i\gamma\omega/q},
\label{gamma}
\eeq
where
$\xi\gg b^{-1}$ is the correlation length 
(we omit the spin symbols for brevity).
 Conventional reasoning for this case (see, e.g., Ref.~\cite{schofield:99}) is that Umklapp scattering is 
 accounted for  
 if $\tau_{\mathrm{ee}}$ is 
  replaced  by the transport time $\tau_{\rm ee}^{{\rm tr}}$.
  For the interaction in Eq.~(\ref{gamma}), $\otau\propto T^2$ for $T\ll T_{\mathrm{FL}}\equiv 1/\gamma \xi^3$ and $\otau\propto T^{D/3}$ for $T\gg T_{\mathrm{FL}}$, while $1/\tau_{\rm ee}^{\rm tr}\sim \left(\otau\right)\left({\bar q}/k_F\right)^2$, where ${\bar q}=\max\{\xi^{-1},(\gamma T)^{1/3}\}$. 
One then obtains  $\rho\propto T^\alpha$,  where $\alpha=2$ in the FL regime 
 and  $\alpha=(D+2)/3$ 
  in the non-FL regime.  In 3D, $\alpha=5/3$  in the  non-FL regime,
 which is close to what has been observed 
  in a number of itinerant ferromagnets \cite{exp_fm}. 

In this Letter we re-examine the role of Umklapp scattering and also 
 analyze the interplay between 
{\em ee} and electron-impurity ({\em ei}) interactions 
 near a Pomeranchuk instability.  We argue that Umklapp processes
 do not give rise to
 the $T^\alpha$ behavior
 of the resistivity near a QPT in a clean system.
  In a dirty system, the {\it correction} to the residual resistivity 
 %does
  scales nominally 
  as $AT^{\alpha}$ at low $T$, but whether  $A$ is finite 
  depends on dimensionality of the FS (2D vs 3D), its topology (simply connected vs multiply connected), and its shape (convex vs concave). 
 For a simply-connected,
 convex FS in 2D, $A=0$, 
 and the 
 first non-vanishing term in $\rho$ scales as $T^{\alpha+2}$,
 which is always subleading to a $T^2$ contribution from 
 non-critical scattering channels. The reason for 
the vanishing of $A$ 
is hidden integrability:
 the constraint that all electrons involved in scattering must be on the FS
lowers the effective dimensionality of scattering events 
from 2D to 1D, where the motion is integrable and hence no relaxation is possible.
%; indeed, 
%(a two-particle collision in 1D can only swap their initial velocities).
 Moving away from the FS breaks integrability but at the price of an extra $T^2$ factor. For a 3D, or multiply-connected, or concave FS, integrability is broken, and $A\neq 0$. However, if these features are weak, i.e., the FS is quasi-2D, there is a crossover between integrable-like scaling (with exponent $\alpha+2$) at higher $T$ to non-integrable--like scaling  (with exponent $\alpha$) at lower $T$. 
We also show 
that 
the resistivity saturates at high $T$, when {\em ee} scattering dominates,
and that a 
 true scaling regime, where $\rho(T)\gg \rho(0)$, can be achieved in a two-band ($s-d$) system with substantially different masses of charge carriers.

The $q=0$ nature of the QPT makes our case different from 
 the one near an antiferromagnetic QPT \cite{rosch}.
 There, Umklapps in both "hot" and "cold" parts of the FS  do lead to finite resistivity, while disorder changes the balance 
 of hot and cold contributions. 
 For the same reason, the interplay between normal and Umklapp processes in a 2D Hubbard model 
 %was studied in Ref.~\
\cite{maebashi97_98} is also different from our case.
% because of the $q=0$ instability. 

%The $q=0$ nature of the instability makes our case different from 
%an interplay between interactions and disorder 
% the one near 
%antiferromagnetic QPT \cite{rosch}. In the latter case, Umklapps in both "hot" and "cold" parts of the FS  do give rise to finite
%resistivity, while disorder changes the balance 
% of hot and cold contributions. The interplay between normal and Umklapp processes in a 2D Hubbard model was studied in Ref.~\cite{maebashi97_98}.  Our case is, again, different because of the $q=0$ instability. 

\begin{figure}[t]
\includegraphics[width=0.4\textwidth]{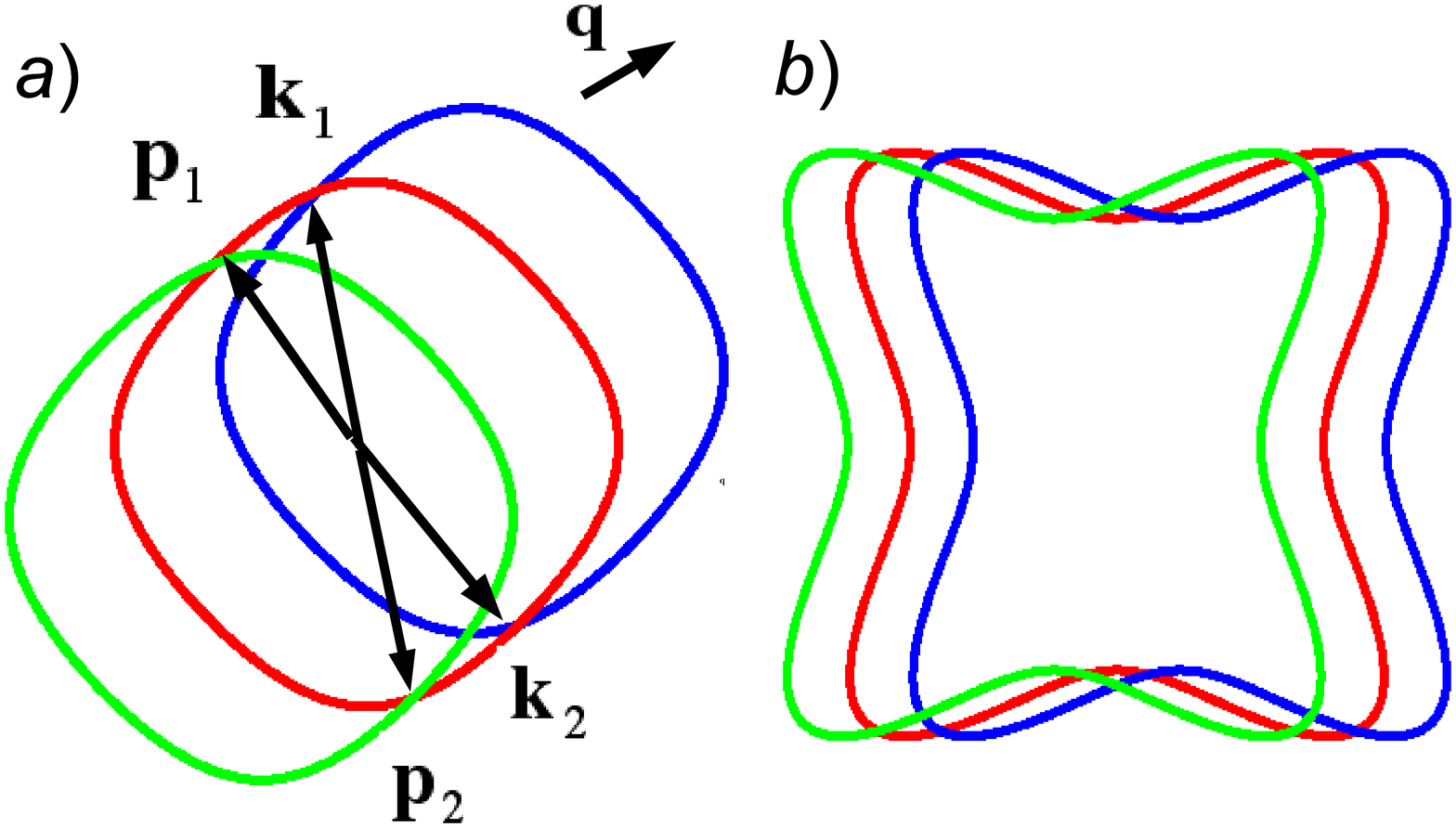}
\caption{(COLOR ONLINE) {\em a})  Normal scattering on a convex, simply connect, 2D Fermi surface. The blue and green FSs are obtained by shifting  the original one (red) by momenta $\bq$ and $-\bq$, respectively.
Process $(\bk_1,\bp_2)\leftrightarrows (\bk_2,\bp_1)$ is a Cooper channel scattering. Processes $(\bk_{1,2},\bp_{1,2})\leftrightarrows (\bp_{1,2},\bk_{1,2} )$ are momentum swaps. Neither of these processes affect the resistivity. {\em b}) 
Mormal scattering on a concave 2D FS. 
 The initial and final states can be chosen from a set of twelve points where the shifted FSs intersect the original one. }  \label{fig:fig2}
\end{figure}

{\it Umklapp scattering.}~~~~
 The
 % conventional 
relation $\rho \propto 1/\tau_{\rm ee}^{\rm tr}$ is based on the 
   assumption that Umklapp events are as frequent as normal ones. 
 We argue that this assumption  breaks down near Pomeranchuk criticality. Indeed,  for  small-angle  scattering,
 one of the final momenta  has to be close to the initial one, e.g., $|{\bf k}-{\bf k}'|\lesssim  {\bar q}\ll b$. This implies that 
 $\bb$ is to be absorbed almost entirely by ${\bf p}-{\bf p'}$, which is only possible if ${\bf p}$ and ${\bf p'}$ are 
 at the edges of the Brillouin zone (and the FS is open), see Fig.~\ref{fig:fig1}. As a result, the Umklapp rate is proportional to the phase space of 
"Umklapp hot spots" and is small by a factor of ${\bar q}^D$ compared to $1/\tau_{\rm ee}^{\rm tr}$ . 
 The conditions
for the 
Umklapp hot spots
 to occur are rather stringent, e.g, they do not exist in a particle-hole symmetric system:
 if Umklapps
 % scattering is 
 are forbidden for a closed FS (at less-than-half filling), they are 
 %it is 
 also forbidden for an open FS (at more-than-half filling). If particle-hole symmetry is broken, 
Umklapp hot spots
 do appear. 
 However, one has to distinguish between  real and pseudo Umklapp processes. For example, the process in Fig.~\ref{fig:fig2}{\em a}
is a pseudo-Umklapp process  because it  can be viewed 
either as an Umklapp event on the open (electron) FS or as a normal event on the closed (hole) FS. 
Since normal scattering does not  give finite resistivity, 
the same is true for 
this type of Umklapp scattering.

To emphasize the difference between the real and pseudo Umklapp processes, we relax the assumption of small $q$ for  a moment. A graphical construction for a closed FS is shown in Fig.~\ref{fig:fig1}{\em b}. If $q$ is larger than some critical value, the Bragg replicas of shifted FSs (magenta and green) intersect the original FS (red) at more than four points. These points represent the initial and final states of Umklapp processes  (shown by arrows), which cannot be mapped onto normal ones. These are real Umklapp processes which do give rise to finite  $\rho$. 
For the case  in Fig.~\ref{fig:fig2}{\em b}, real Umklapps occur if
$b-k_{\max}<q_x<b$, where $k_{\max}$ is the maximal diameter of the FS in the $x$ direction. For small $q$, this can happen only near half-filling, when $|k_{\max}-b|\lesssim {\bar q}$. However, 
 half-filling favors a finite-$q$ instability, e.g., antiferromagnetism, over the $q=0$ one. Away from half-filling, real umklapp processes can only happen for ${\bar q}\sim b$,
due to the non-critical part of the interaction. This gives rise to a conventional $T^2$ term in $\rho$.

 The conclusion of this analysis is that umklapp scattering in a clean system
 with small-angle scattering cannot give rise to a critical behavior $\rho(T)
\propto T^{\beta}$ with $\beta<2$.

{\it A combination of normal ee and ei interactions.}~~~
  We now neglect Umklapp processes but invoke impurity scattering as a mechanism
 of momentum relaxation. Our analysis is based on the Boltzmann equation near equilibrium (its validity is discussed later in this Letter) 
\beq
e{\bf v}_{\bk}\cdot {\bf E}n'_{\bk}=-I_{\mathrm{ei}}-I_{\mathrm{ee}},\label{be}
\eeq
where ${\bf E}$ is the electric field,
  $n_{\bk}$ is the Fermi funtion,
  and  $I_{\mathrm{ei}}$ and $I_{\mathrm{ee}}$ describe the {\em ei} and {\em ee} scatterings.
Although all of our results are valid for the most general form of 
$I_{\mathrm{ei}}$, we will restrict our attention to $\delta$ function impurities, when $I_{\mathrm{ei}}=\left(f_{\bk}- n_{\bk}\right)/\ti$ with $\ti=\mathrm{const}$. The {\em ee} collision integral for the non-equilibrium part of $f_{\bk}$ defined by $f_{\mathbf{k}}=
%n_{\mathbf{k}}-Tn_{\mathbf{k}}^{\prime }g_{\mathbf{k}}=
n_{\mathbf{k}}+n_{\mathbf{k}}\left(1-n_{\mathbf{k}}\right) g_{\mathbf{k}}$ can be written as \cite{abrikosov}
\begin{eqnarray}
&&I_{\mathrm{ee}} =\sum_{\bp,\bq}  |M_{\mathbf{k,p}}(\bq,\epsilon_{\bk}-\epsilon_{\bk-\bq})|^2 \left( g_{\mathbf{k}}+g_{%
\mathbf{p}}-g_{\mathbf{k}-\bq}-g_{\mathbf{p}+\bq}\right)\notag\\
&&\!\!\!\times n_{%
\mathbf{k}}n_{\mathbf{p}}\left( 1-n_{\mathbf{k}-\bq}\right) \!\left(
1-n_{\mathbf{p}+\bq}\right) 
\!\delta \left( \epsilon _{\mathbf{k}}+\epsilon _{\mathbf{p}%
}-\epsilon _{\mathbf{k}-\bq}-\epsilon _{\mathbf{p}+\bq}\right). \label{iee} 
\end{eqnarray}
where $M_{\bk,\bp}(\bq,\omega) 
% \propto 
=U_{\mathrm{eff}} (\bq, \omega)S_{\bk,\bp}$
  is the matrix element  of the  effective {\em ee} interaction
  on the Bloch wave functions, and $S_{\bk,\bp}$ is the structure factor for a given lattice.  Normal {\em ee} collisions  conserve 
  the momentum,  i.e.,  $\sum_{\bk}\bk I_{\mathrm{ee}}=0$.   For a GIFL with ${\bf v}_{\bk}=\bk/m$, the conductivity is obtained by multiplying Eq.~(\ref{be})  by $\bv_{\bk}$ and summing over $\bk$, upon which $I_{\mathrm{ee}}$ drops out, so that the resulting relation between the electrical current and ${\bf E}$ is independent of the {\em ee} interaction. For a non-GIFL with ${\bf v}_{\bk}={\bf\nabla}_{\bk}\epsilon_{\bk}\neq\bk/m$, normal collisions, in general, affect the conductivity. 

{\it Low temperatures.}~~~
The first question   
is whether the correction to the residual 
conductivity due to normal {\em ee} scattering
 scales as $AT^{\alpha}$ at low $T$, 
 when $\tau_{\mathrm{ee}}\gg \ti$.  Solving Eq.~(\ref{be}) to first order in $I_{\mathrm{ee}}$, we obtain
 \begin{widetext}
\begin{eqnarray}
\delta\sigma _{ii} &=&-\frac{e^{2}\ti^2}{2T}\int \frac{d^{D}q}{\left( 2\pi
\right) ^{D}}\int \int \int d\omega d\epsilon _{\mathbf{k}}d\epsilon _{%
\mathbf{p}}\oint \oint \frac{da_{\bk}}{ v_{\bk} }%
\frac{da_{\bp}}{v_{\bp}}|M_{\mathbf{k,p}}\left( 
\mathbf{q},\omega\right)|^2
\Delta\bv^2_i
 \notag \\
&&\times n\left( \epsilon _{\mathbf{k}}\right) n\left(
\epsilon _{\mathbf{p}}\right) \left[ 1-n\left( \epsilon _{%
\mathbf{k}}-\omega \right) \right] \left[ 1-n\left( \epsilon
_{\mathbf{p}}+\omega \right) \right] \delta \left( \epsilon _{\mathbf{k}%
}-\epsilon _{\mathbf{k-q}}-\omega \right) \delta \left( \epsilon _{%
\mathbf{p}}-\epsilon _{\mathbf{p}+\mathbf{q}}+\omega \right), 
\label{sigma_gen}
\end{eqnarray}
\end{widetext}
where 
 $\Delta\bv\equiv {\bf v}_{\bk}+{\bf v}_{\bp}-{\bf v}_{\bk-{\bf q}}-{\bf v}_{\bp+{\bf q}}$,
and $da_{\bf l}$ is the FS element.
  For a GIFL, $\Delta\bv=0$ and thus $\delta\sigma_{ii}$ vanishes identically.  We will see, however, that the leading term in $\delta\sigma_{ii}$ may vanish  even on a lattice. The crucial point is that the leading $T$ dependence of $\delta\sigma_{ii}$ is obtained by neglecting $\omega$ in both $\delta$ functions, i.e., by projecting electrons onto the FS.  Integrating over $\epsilon_{\bk}$ and $\epsilon_{\bp}$, we obtain 
 \begin{eqnarray}
&&\delta\sigma _{ii } =-\frac{e^{2}\ti^2T^2}{2}\int \frac{d^{D}q}{\left( 2\pi
\right) ^{D}}\oint \oint \frac{da_{\bk}}{ v_{\bk} }%
\frac{da_{\bp}}{v_{\bp}}
R_{\mathbf{k,p}}
\left( \mathbf{q}\right) \notag\\
 &&\times
 \Delta\bv_i^2 \delta \left( \epsilon _{\mathbf{k}}-\epsilon _{\mathbf{k-q}}\right)\vert_{\epsilon_{\bk}=0}
 \delta \left( \epsilon _{
\mathbf{p}}-\epsilon _{\mathbf{p}+\mathbf{q}} \right)\vert_{\epsilon_{\bp}=0}
, 
\label{sigma_gen_1}
\end{eqnarray}
 where $R_{\mathbf{k,p}}\left(\mathbf{q}\right)\equiv \int d\omega\left(\omega^2/T^3\right)|M_{\mathbf{k,p}}\left( 
\mathbf{q},\omega\right)|^2N(\omega)\\
\times\left[N(\omega)+1\right]$ and $N(\omega)$ is the Bose function.  
  By power counting, 
$\delta\sigma_{ii} \propto AT^{\alpha}$;
 yet one has to verify if  $A \neq 0$.

The prefactor $A$  is given by the solution of a purely geometric problem: for a fixed  momentum transfer $\bq$, find the initial states $\bk$ and $\bp$ on the FS so that energy conservation is satisfied. For small $q$, energy conservation reduces to
$\epsilon_{\bk}-\epsilon_{\bk-\bq}\approx \bv_{\bk}\cdot\bq=0$ and $\epsilon_{\bp}-\epsilon_{\bp+\bq}\approx -\bv_{\bp}\cdot\bq=0$.
Therefore, $\bk$ and $\bp$ are the points where {\em all} tangents in the direction of $\bq$ intersect the FS. Since a convex, simply-connected, and 2D FS has only two tangents, there are only two solutions for $\bk$ and $\bp$ \cite{maebashi97_98}. Equivalently,  if the FS of this type is shifted by a small momentum $\bq$, there are only two intersection points ($\bk_1$ and $\bk_2$ in Fig.~\ref{fig:fig2}{\em a}). A shift by $-\bq$  gives two more points: $\bp_1$ and $\bp_2$.  
 However, these solutions are not independent. Indeed, since the equation $\epsilon_{\bk-\bq}=\epsilon_{\bk}$ has only two roots, and the second equation $\epsilon_{\bp+\bq}=\epsilon_{\bp}$  reduces to the first one upon
 $\bp\to-\bk$, we have $\bp_1=-\bk_2,\bp_2=-\bk_1$. Therefore, the process $(\bk_1,\bp_2)\leftrightarrows(\bp_1,\bk_2)$ corresponds to the Cooper channel of scattering with  zero total momentum.
 But this implies that 
 $\bv_{\bk_1}+\bv_{\bp_2}=\bv_{\bk_2}+\bv_{\bp_1}=0$, 
so that  $\Delta\bv=0$ and $A=0$. In addition, since $-\bk+\bq$ also solves $\epsilon_{\bk-\bq}=\epsilon_{\bk}$, 
 we must have $\bk_{1}=
 %$ must coincide with $
 -\bk_{2}+\bq$ (or v.v.), which implies that $\bp_{a}+\bq=\bk_{a}$ ($a=1,2$).
 Therefore, the remaining process, $(\bk_{a},\bp_{a})\leftrightarrows(\bp_{a},\bk_{a})$, just swaps the initial and final states, and $\Delta\bv=0$ again. Therefore, $A=0$ even though Galilean invariance is broken.

The first non-vanishing term in 
 $\delta\sigma_{ii}$ is obtained by considering electrons slightly away from the FS, i.e.,  by expanding the product of the energy $\delta$ functions to second order in $\omega$. The derivatives of the $\delta$ functions produce the same solutions for $\bk$ and $\bp$ as the $\delta$ functions themselves. 
These solutions nullify $\Delta\bv$ but not its derivatives generated by integration by parts. As a result, $\delta\sigma_{ii}$ is finite but contains an extra factor of $T^2$ compared to the power-counting estimate, i.e., the \lq\lq critical\rq\rq\/ contribution to the resistivity behaves as 
 $\rho_{ii}(T)-\rho_{ii}(0)=BT^{\alpha+2}$.  
 Because $\alpha +2 >2$, the ``critical'' contribution is subleading to a $T^2$ contribution from non-critical channels, e.g., a charge channel near a magnetic instability.
 
A FS of any other type (3D, multiply connected, concave) has more than two self-intersection points when shifted by a small momentum, so that each of the equations $\epsilon_{\bk}=\epsilon_{\bk\pm\bq}$ has more than two solutions: 
infinite number for a 3D FS and finite but larger than two number  for a multiply connected
or concave 2D FS. [cf. Fig.~\ref{fig:fig2}{\em b}]. 
 Thus integrability is broken, and $A\neq 0$. However, in a number of situations (quasi-2D or slightly concave FS), integrability is broken only weakly. 
Suppose, for example, that $\epsilon_{\bk}=\epsilon_{xy}(k_x,k_y)+\epsilon_{z}(k_z)$, where $\epsilon_{xy}(k_x,k_y)=0$ describes a simply connected, convex, and 2D FS,  and $\epsilon_{z}(k_z)=t_{\perp}\left[(1-\cos(k_zc)\right]$ with $t_{\perp}\ll \epsilon_{F}$. The $\delta$ functions can be expanded to second order in both $\omega$ and $\epsilon_{z}$, which produces two types of terms: $\rho_{ii}(T)-\rho_{ii}(0)=B_1T^{\alpha+2}+B_2t_{\perp}^2T^{\alpha}$. For $T\gg t_{\perp}$ ($T\ll t_{\perp}$), the first (second)
% (integrable) 
term dominates.
% the 
%scaling.
 %; for $T\ll t_{\perp}$, the second (non-integrable) term dominates. 

{\it High temperatures}~~~ We now 
show that the resistivity saturates in  the opposite limit of high temperatures, when $\tau_{\mathrm{ee}}\ll \ti$. 
 The proof is similar to the one for normal phonon-phonon collisions \cite{physkin}. Frequent normal {\em ee} collisions establish a quasi-equilibrium distribution 
 with the drift velocity ${\bf u}$,  fixed by rare {\em ei} collisions. 
 Accordingly,  $f_{\bk}=n'_{\bk}\bk\cdot{\bf u}+f_{\bk}^{\{\mathrm{i}\}}$, where the first term nullifies $I_{\mathrm{ee}}$, and the second term is small.
 To first order in $1/\ti$, the Boltzmann equation reads $e\bv_{\bk}\cdot{\bf E}n'_{\bk}=-I_{\mathrm{ee}}[f_{\bk}^{\{\mathrm{i}\}}]-n'_{\bk}\bk\cdot{\bf u}/\ti.$
 Applying $\sum_{\bk}\bk$, we eliminate $I_{\mathrm{ee}}$ and solve for ${\bf u}$ in terms of ${\bf E}$. The current is determined primarily by the first term in $f_{\bk}$, which is independent of the {\em ee} interaction.  Hence 
 the critical
 component
 of $\rho (T)$ saturates at high $T$.
The low- and high-$T$ limits  differ only in how the averaging over the FS is performed: 
$\sigma_{ij}(\infty) = e^2\nu_F \ti
\sum_{l}
\langle v_{i}k_{l}\rangle
\langle v_{j}k_{l}\rangle/\langle k^2_{l}\rangle$,  whereas $\sigma_{ij}(0) = e^2\nu_F\ti\langle v_{i}v_{j}\rangle$ ($\nu_F$ is the density of states). For a  generic case, $\sigma_{ij}(\infty)\lesssim\sigma_{ij}(0)$.
This implies that $\rho_{ii}(T)$ cannot be much larger than $\rho_{ii}(0)$, i.e.,  there is no true scaling regime. If, however,  $\sigma_{ij}(\infty)\ll\sigma_{ij}(0)$, scaling exists in a wide $T$ interval where $\rho_{ii}(0)\ll\rho_{ii}(T)\ll\rho_{ii}(\infty)$. 

{\it  $s-d$ model near criticality.}~~~~  Finally, we show that
 a true $T^{\alpha}$ 
 scaling of the resistivity  near a QPT is possible in a dirty two-band metal
with substantially different band masses ($s-d$ model \cite{sd}).
  The heavy ($d$) band is assumed to be near criticality, the light ($s$) band is not critical on its own, but the interband interaction becomes critical due to renormalization in the $s-d$ channel: $V_{sd}(\bq,\omega)=V_{sd}^{0}/(1-\chi_{dd}(\bq,\omega)V^0_{dd})$, where $\chi_{dd}(\bq\to 0,0)V^0_{dd}\approx 1$. 
 In the absence of Umklapps, we still need to couple each of the bands to impurities  to render $\rho$ finite. The electron-impurity times are such that
  $\tau_{\mathrm{i}s}\propto m_s^{-1}\gg \tau_{\mathrm{i}d}\propto m_d^{-1}$. Since a two-band FS is already non-integrable, we adopt the simplest model of two parabolic bands in 2D  and neglect all other interactions except for $V_{sd} ({\bf q}, \omega)$.
 An exact solution of two coupled Boltzmann equations gives 
\beq
\rho(T)=\frac{\pi\hbar}{e^2\epsilon_F}\frac{\frac{1}{\tau_{\mathrm{i}s}\tau_{\mathrm{i}d}}
+\frac{1}{\tau_{sd}(T)}\left(\frac{1}{\tau_{\mathrm{i}s}}\frac{m_s}{m_d}+\frac{1}{\tau_{\mathrm{i}d}}\frac{m_d}{m_s}\right) }{\frac{1}{\tau_{\mathrm{i}s}} +\frac{1}{\tau_{\mathrm{i}d}}+\frac{1}{\tau_{sd}(T)}\left(2+\frac{m_s}{m_d}+\frac{m_d}{m_s}\right)},
\label{2band}\eeq
where 
${\tau_{sd}^{-1}(T)}=\frac{(m_s m_d)^{1/2}}{2T\epsilon_F^2}\!\!\!\int\!\!\!\int d\omega dqq |V_{sd}
%(\bq,\omega
|^2\omega^2N(\omega)\left[N(\omega)+1\right].$
At criticality, $1/\tau_{sd}(T)\propto T^{4/3}$. The low- and high-$T$ limits are controlled by the $s$ and $d$ electrons, respectively: $\rho(0)\approx \pi\hbar/e^2\epsilon_F\tau_{\mathrm{i}s}\ll \rho(\infty)\approx \pi\hbar/e^2\epsilon_F\tau_{\mathrm{i}d}$. The scaling regime corresponds to the interval  
$T_1\ll T\ll T_2$, where $\tau_{sd}(T_1)=\tau_{\mathrm{i}s}m_d/m_s$ and $\tau_{sd}(T_2)=\tau_{\mathrm{i}d}m_d/m_s$.  In this  regime, $\rho(T)$ is independent of
disorder and behaves as
 $\rho(T)=\left(\pi\hbar/e^2\right)\left(m_d/m_s\right)\left(1/\epsilon_F\tau_{sd}\right)\propto T^{4/3}$ (similarly, $\rho\propto T^{5/3}$ in 3D). Since quantum-critical metals typically have light and heavy bands, 
 it is quite possible  that  the $s-d$ physics 
 is responsible for the observed  critical 
 scaling of the resistivity.
Equation~(\ref{2band}) 
also applies to a ferromagnetic metal with only band in the paramagnetic phase. 
In this case, ``$s$'' and ``$d$'' refer to spin-up and spin-down electrons.
The $T^{(D+2)/3}$ term in $\rho$ is, however, non-zero only in the symmetry-broken phase.

{\it Limitations of the Boltzmann-equation approach.}~~~
An obvious deficiency of the semiclassical Boltzmann equation is that
 it neglects both quantum \cite{altshuler:85} and viscous 
 \cite{spivak:02} corrections to resistivity. 
Both effects are, in general, relevant but,
 in wide $T$ intervals, 
they are smaller than the direct contribution of the {\em ee} interaction to the resistivity, 
$\delta\rho_{\mathrm{d}}$
discussed in this paper, if the latter is not suppressed by integrability.
   The quantum-interaction correction  
$\delta\rho_{\mathrm {QI}}$ is smaller than $\delta\rho_{\mathrm{d}}$ 
in the ballistic regime, where ${\bar q}v_F\ti\gg1$:
 in a 2D FL,  $|\delta\rho_{\mathrm {QI}}|/\rho(0)\sim T/\epsilon_F$ \cite{zala:01} while $|\delta\rho_{{\mathrm d}}|/\rho(0)\sim T^2\ti/\epsilon_F$, so that 
 $\delta\rho_{{\mathrm d}}/|\delta\rho_{{\mathrm QI}}|\sim T\ti \gg 1$;
 the same is true also in the non-FL regime \cite{paul:05}.   
  In the diffusive regime (where ${\bar q}v_F\ti\ll1$), $|\delta\rho_{\mathrm{ QI}}|\gg \delta\rho_{d}$.
 The viscous correction is also
 smaller than $\delta\rho_{\mathrm{d}}$, 
 if the impurity scattering length is smaller than $v_F\ti$.

  The   Boltzmann 
 approach may also fail because quasiparticles
 are not well-defined  in the non-FL regime.  
 However, if the critical {\em ee} interaction can be treated within the Eliashberg approximation, the validity of the Boltzmann equation does not rely on the assumption  of well-defined quasiparticles -- the proof follows  the Prange-Kadanoff  reasoning  for an electron-phonon system~\cite{prange:64}.
  Although recent findings \cite{eliashberg_breakdown} indicate that the Eliashberg approximation 
 for the self-energy
 is not controlled for $D=2,Z=3$ criticality, it is  possible that transport properties,
 which are less sensitive to infrared singularities,
 can
 still be 
 described  
 within 
 this approximation.

As a final remark,  
 we note that some of our results are applicable beyond the model with interaction in Eq.~(\ref{gamma}). In particular, all results for the FL regime do not depend on a particular 
 form
 of the interaction, as long as it remains long-ranged. Moreover, 
 integrability exists for any interaction
 on a small yet anisotropic Fermi surface.

We thank A. Kamenev, D. Loss, H. Pal, I. Paul, C. P{\'e}pin, M. Reizer, S. Sachdev, B. Spivak, and A. Varlamov for interesting discussions.
This work was supported by NSF-DMR-0908029 (D.L.M.),  RFBR-09-02-01235 (V.I.Y.), and NSF-DMR-0906953
 (A.V.Ch.). D.L.M. and A.V.Ch. acknowledge hospitality of MPI-PKS (Dresden), where a part of this work was done.

\end{document}